\documentclass[aps,showpacs,amsmath,preprint,amssymb,12pt]{revtex4}
\usepackage{graphicx}
\begin{document}
{~}
\vspace{3cm}

\title{Charged Rotating Kaluza-Klein Black Holes in Five Dimensions
\vspace{1cm}
}
\author{Toshiharu Nakagawa, Hideki Ishihara, Ken Matsuno, and Shinya Tomizawa}
\affiliation{ 
Department of Mathematics and Physics,
Graduate School of Science, Osaka City University,
3-3-138 Sugimoto, Sumiyoshi, Osaka 558-8585, Japan
\vspace{3cm}
}

\begin{abstract}
We construct a new charged rotating Kaluza-Klein black hole solution in the five-dimensional Einstein-Maxwell theory with a Chern-Simon term. 
The features of the solutions are also investigated. The spacetime is asymptotically locally flat, i.e., it asymptotes to a twisted $\rm S^1$ bundle over the four-dimensional Minkowski spacetime. 
The solution describe a non-BPS black hole rotating in the direction of the extra dimension. The solutions have the limits to the supersymmetric black hole solutions, a new extreme non-BPS black hole solutions and a new rotating non-BPS black hole solution with a constant twisted $\rm S^1$ fiber.
\end{abstract}

\date{\today}

\preprint{OCU-PHYS 286}
\preprint{AP-GR 52}

\maketitle

\section{Introduction}
In the context of string theory, the five-dimensional Einstein-Maxwell theory
with a Chern-Simon term gathers much attention since it is the bosonic sector of the minimal supergravity. 
Supersymmetric (BPS) black hole solutions to the five-dimensional Einstein-Maxwell 
equations with a Chern-Simon term have been found by various authors. 
Based on the classification of the five-dimensional supersymmetric solutions 
by Gauntlett.et.al.~\cite{Gauntlett0}, they have been constructed on hyper-K\"ahler 
base spaces, especially, the Gibbons-Hawking base space. 
The first asymptotically flat supersymmetric black hole solution, BMPV (Breckenridge-Myers-Peet-Vafa) solution, was constructed on the four-dimensional Euclid space~\cite{BMPV}. 
A supersymmetric black hole solution with a compactified extra dimension 
on the Euclidean self-dual Taub-NUT base space
was constructed by Gaiotto.et.al~\cite{Gaiotto}. 
It was extended to multi-black hole solution with the same asymptotic structure~\cite{IKMT}.
One of the most interesting properties is that the possible spatial topology of the horizon of each black hole is the lens space S$^3/{\mathbb Z_{n_i}}$ in addition 
to $\rm S^3$, where $n_i$ are the natural numbers . 
Similarly, black hole solution on the Eguchi-Hanson space~\cite{IKMT2}, which admits horizon topologies of various lens spaces $L(2n;1)={\rm S}^3/{\mathbb Z}_{2n}$ ($n$:natural number), was also constructed. 
Since the latter black hole spacetimes are asymptotically locally Euclidean, we cannot locally distinguish these asymptotic structure. 

As solutions in five-dimensional Einstein-Maxwell theory with a positive cosmological constant, black hole solutions 
on the Euclid base space~\cite{London}, the Taub-NUT base space~\cite{IIKMMT} and 
the Eguchi-Hanson base space~\cite{IMK} were also constructed. 
In particular, two-black hole solution on the Eguchi-Hanson space describes 
a non-trivial coalescence of black holes. 
In refes. \cite{IMK,YIKMT}, the authors compared the two-black hole solution 
on the Eguchi-Hanson space with the two-black holes solution 
on the Euclid space~\cite{London}, 
and discussed how the coalescence of five-dimensional black holes depends on 
the asymptotic structure of spacetime. 
Two black holes with the topology of $\rm S^3$ coalesce into a single black hole with the topology of the lens space $L(2;1)=\rm S^3/{\mathbb Z}_2$ in the case of 
Eguchi-Hanson space, while two black holes with the topology of $\rm S^3$ coalesce 
into a single black hole with the topology of $\rm S^3$ in the Euclid case. 
When the action has the Chern-Simon term, black holes can rotate~\cite{KS,MIKT}.

The most remarkable feature of the five-dimensional black hole solutions is that
they admit non-spherical topology. Emparan and Reall found the first black ring 
solution to the five-dimensional vacuum Einstein equations~\cite{Emparan}, 
which describes a stationary, rotating black hole with the horizon topology 
$\rm S^1\times S^2$. 
Some supersymmetric black ring solutions have been also found, based on the 
construction of the solutions by Gauntlett.et.al.~\cite{Gauntlett0}.  
Elvang et.al. found the first supersymmetric black ring solution with asymptotic 
flatness on the four-dimensional Euclidean base space, which is specified by three 
parameters, mass and two independent angular momentum components~\cite{Elvang}. 
Gauntlett and Gutowski also constructed a multi-black ring solution on the same 
base space~\cite{Gauntlett,Gauntlett2}. 
The BPS black rings with three arbitrary charges and three dipole charges on the flat space were also constructed in ~\cite{Elvang2,Bena3}. 
The BPS black ring solutions on the Taub-NUT base 
space~\cite{Bena,Bena2,Gaiotto2,Elvang2} and the Eguchi-Hanson space~\cite{TIKM}were constructed. 
The black ring solution on Eguchi-Hanson space has the same two angular momentum components and the asymptotic structure on time slices is asymptotically locally Euclidean. The $\rm S^1$-direction of the black ring is along the equator on a $\rm S^2$-bolt on the Eguchi-Hanson space. 
It has the limit to the BMPV black hole with the topology of the lens space 
$\rm L(2;1)=S^3/Z_2$.

In addition to the BPS solutions, 
the non-BPS black hole solutions have also been studied by several authors. 
Cvetic et.al.~\cite{CLP} found a non-extremal, charged  and rotating black hole solution with asymptotic flatness.~\footnote{We consider the case of vanishing cosmological constant.} 
In the specified limits, the solution reduces to the known solutions: 
the same angular momenta case of the Myers-Perry black hole solution~\cite{Myers}, Exact solutions of non-BPS Kaluza-Klein black hole solutions are found in 
neutral case~\cite{DM,GW} and charged case~\cite{IM} . 
These solutions have a non-trivial asymptotic structure, i.e., 
they asymptotically approach a twisted $\rm S^1$ bundle over the four-dimensional 
Minkowski spacetime. The horizons are deformed due to this non-trivial asymptotic 
structure and have a shape of a squashed $\rm S^3$, where $\rm S^3$ is regarded as 
a twisted bundle over a $\rm S^2$ base space. The ratio of the radius $\rm S^2$ to 
that of $\rm S^1$ is always larger than one.

 Wang proposed that a kind of Kaluza-Klein black hole solutions 
can be generated by the \lq {\it squashing transformation}\rq~ from 
black holes with asymptotic flatness~\cite{Wang}. 
In fact, he regenerated the five-dimensional Kaluza-Klein black hole solution found 
by Dobiasch and Maison~\cite{DM,GW}
from the five-dimensional Myers-Perry black hole solution with two equal angular momenta.\footnote{
The solution generated by Wang coincides with the solution in Ref.\cite{DM,GW}. }

In this article, using the squashing transformation, we construct a new non-BPS rotating charged Kaluza-Klein black hole solutions in the five-dimensional 
Einstein-Maxwell theory with a Chern-Simon term, which is the generalization of the Kaluza-Klein black hole solution in Ref.~\cite{DM,GW,IM}. 
Applying the squashing transformation to the Cvetic et.al.'s charged rotating black hole solution~\cite{CLP} 
in vanishing cosmological constant case, 
we obtain the new Kaluza-Klein black hole solution. 
We also investigate the features of the solution. The spacetime is asymptotically locally flat, i.e., it asymptotically approaches a twisted $\rm S^1$ bundle over the four-dimensional Minkowski spacetime. 
The solution describes a non-BPS black hole boosted in the direction of the extra dimension. It has various limits, f.g., to the supersymmetric BMPV black hole solution, 
to the supersymmetric Kaluza-Klein black hole solution 
and to the extreme non-BPS black hole solutions.

The rest of this article is organized as follows. 
In Sec.\ref{sec:solution}, we present a new Kaluza-Klein black hole solution in the five-dimensional Einstein-Maxwell theory with a Chern-Simon term. 
In Sec.\ref{sec:feature}, we investigate the basic features of the solution. In Sec.\ref{sec:limit}, we study the limit of our solution to various known and unknown black hole solutions. In Sec.\ref{sec:summary}, we summarize the results in this article.

\section{Solution}\label{sec:solution}
In the metric of squashing Kaluza-Klein black hole in ref.~\cite{IM}, 
a function of the radial coordinate $k(r)$, which describes the squashing of the 
horizons, appeares. Wang pointed out that the function $k(r)$ would give a 
transformation from asymptotically flat solutions to Kaluza-Klein type solutions.
He call this {\it squashing transformation. 
}

\begin{figure}[htbp]
\begin{center}
\includegraphics[width=0.6\linewidth]{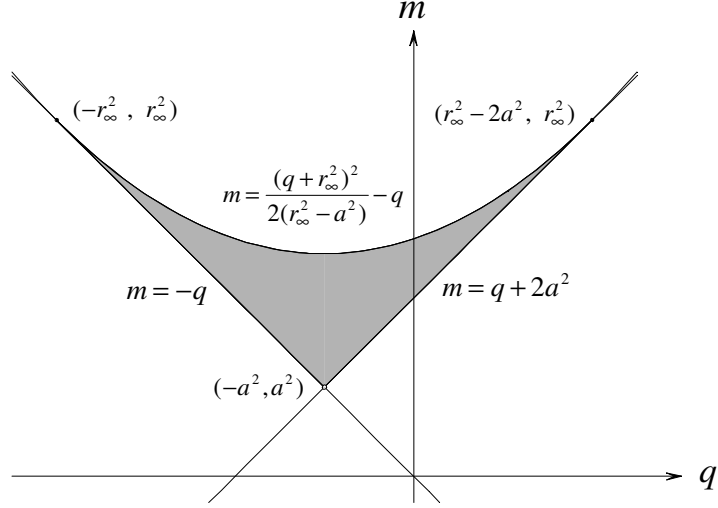}
\end{center}
\caption{Region of the parameters in the $(q,m)$-plane.
}
\label{fig:region}
\end{figure}

Applying the squashing transformation to the non-BPS charged rotating black hole 
solution found by Cvetic et.al.~\cite{CLP}, we construct a new charged, rotating 
Kaluza-Klein black hole solution to the five-dimensional Einstein-Maxwell theory 
with a Chern-Simon term. 
The metric and the gauge potential of the solution are given by
\begin{eqnarray}
ds^2&=&-\frac{w(r)}{h(r)}dt^2+k(r)^2\frac{dr^2}{w(r)}+\frac{r^2}{4}\biggl[k(r)(\sigma ^2_{1}+\sigma ^2_2)+h(r)(f(r)dt+\sigma _3)^2\biggr],
\end{eqnarray}
and 
\begin{eqnarray}
A=\frac{\sqrt{3}q}{2r^2}\left(dt-\frac{a}{2}\sigma _3\right),
\end{eqnarray}
respectively, where the metric functions $w(r), h(r),f(r)$ and $k(r)$ are defined as
\begin{eqnarray}
&&w(r)=\frac{(r^2+q)^2-2(m+q)(r^2-a^2)}{r^4},\\
&&h(r)=1-\frac{a^2q^2}{r^6}+\frac{2a^2(m+q)}{r^4},\\
&&f(r)=-\frac{2a}{r^2h(r)}\left(\frac{2m+q}{r^2}-\frac{q^2}{r^4}\right),\\
&&k(r)=\frac{(r_\infty^2+q)^2-2(m+q)(r_\infty^2-a^2)}{(r_\infty^2-r^2)^2},
\end{eqnarray}
and the left-invariant $1$-forms on $\rm S^3$ are given by
\begin{eqnarray}
&&\sigma_1=\cos\psi d\theta+\sin\psi\sin\theta d\phi,\\
&&\sigma_2=-\sin\psi d\theta+\cos\psi\sin\theta d\phi,\\
&&\sigma_3=d\psi+\cos\theta d\phi.
\end{eqnarray} 
The coordinates $r,\theta,\phi$ and $\psi$ run the ranges of $0<r<r_\infty$, $0\le \theta<\pi$, $0\le \phi<2\pi$, $0\le \psi<4\pi$, respectively. In the case of $k(r)=1$, i.e., $r_\infty\to\infty$, the metric coincides with that of the Cvetic et.al.'s solution without a cosmological constant. 
In this article we assume that the parameters $a,m,q,$ and $r_\infty$ appearing in the solutions satisfy the inequalities
\begin{eqnarray}
&&m>0,\label{eq:0in2}\\
&&q^2+2(m+q)a^2>0,\label{eq:0in3}\\
&&(r_\infty^2+q)^2-2(m+q)(r_\infty^2-a^2)>0,\label{eq:0in3-2}\\
&&(m+q)(m-q-2a^2)>0\label{eq:0in4},\\
&&m+q>0,\label{eq:0in1}.
\end{eqnarray}
As will be explained later, the inequalities (\ref{eq:0in2})-(\ref{eq:0in4}) are the conditions for the existence of two horizons, and the condition (\ref{eq:0in1}) is the requirement for the absence of closed timelike curves outside the outer horizon. 
FIG.\ref{fig:region} shows the region of the parameters.

\section{Features of the solutions}\label{sec:feature}
\subsection{Asymptotic form}
In the coordinate system $(t,r,\theta,\phi,\psi)$, the metric diverges at $r=r_\infty$ but we see that this is an apparent singularity and corresponds to the spatial infinity. To confirm this, introduce a new coordinate defined by
\begin{eqnarray}
\rho=\rho_0\frac{r^2}{r_\infty^2-r^2},
\end{eqnarray}
where the constant $\rho_0$ is given by
\begin{eqnarray}
\rho_0^2=\frac{(r^2_\infty+q)^2-2(m+q)(r_\infty^2-a^2)}{4r_\infty^2}.\label{eq:rho0}
\end{eqnarray}
In terms of the coordinate $\rho$, the metric is written as follows
\begin{eqnarray}
ds^2=-V(\rho)dt^2+2H(\rho)dt\sigma_3+U(\rho)d\rho^2+\rho(\rho+\rho_0)(\sigma_1^2+\sigma_2^2)+W(\rho)\sigma_3^2,
\end{eqnarray}
where the functions $V(\rho),\ H(\rho),\ U(\rho)$ and $W(\rho)$ are given by
\begin{eqnarray}
&&V(\rho )=\frac{\left((r^2_{\infty }+q)\rho +\rho_{0}q \right)^2-2(m+q)r^2_{\infty }\rho(\rho+\rho_{0})}{r^4_{\infty }\rho^2}, \label{eq:f1}\\
&&H(\rho)=\frac{a(\rho+\rho_{0})\left((q^2-(2m+q)r_{\infty }^2))\rho+q^2\rho_{0}\right)}{2r_{\infty }^4\rho^2},\\
&&U(\rho)=\frac{4r^2_{\infty }\rho_{0}^2 \rho(\rho+\rho_{0})}{(q^2+2a^2(m+q))(\rho+\rho_{0})^2-2mr_{\infty }^2\rho(\rho+\rho_{0})+r_{\infty }^4\rho^2}, \\
&&W(\rho)=\frac{r_{\infty }^6\rho^3-a^2(\rho+\rho_{0})^2\left((q^2-2(m+q)r_{\infty }^2)\rho+q^2\rho_{0}\right)}{4r_{\infty }^4\rho^2(\rho+\rho_{0})}.\label{eq:f2}
\end{eqnarray}

This new radial coordinate $\rho$ runs from $0$ into $\infty$.
For $\rho\to \infty$, which corresponds to the limit of $r\to r_\infty$, the metric behaves as
\begin{eqnarray}
ds^2&\simeq& -\frac{(r^2_{\infty}+q)^2-2(m+q)r^2_{\infty}}{r_\infty^4}dt^2+\frac{a(q^2-(2m+q)r_\infty^2)}{r_\infty^4}dt\sigma_3\\
     & &+d\rho^2+\rho^2(\sigma_1^2+\sigma_2^2)+\frac{r_\infty^6-a^2(q^2-2(m+q)r_\infty^2)}{4r_\infty^4}\sigma_3^2.
\end{eqnarray}
Next in order to transform the asymptotic frame into the rest frame, we define the coordinates $(\tilde t,\tilde\psi)$ given by
\begin{eqnarray}
\tilde \psi&=&\psi+\frac{2a(q^2-(2m+q)r_\infty^2)}{r_\infty^6-a^2(q^2-2(m+q)r_\infty^2)}\ t,\\
\tilde t&=&\sqrt{\frac{4r_\infty^4\rho_0^2}{r_\infty^6-a^2(q^2-2(m+q)r_\infty^2)}}\ t.
\end{eqnarray}
Then the metric takes the following asymptotic form
\begin{eqnarray}
ds^2\simeq -d\tilde t^2+d\rho^2+\rho^2(\sigma_1^2+\sigma_2^2)+L^2\tilde\sigma_3^2,
\end{eqnarray}
where $\tilde \sigma_3=d\tilde\psi+\cos\theta d\phi$ and the size of extra dimension $L$ is given by
\begin{eqnarray}
L^2=\frac{r_\infty^6-a^2(q^2-2(m+q)r_\infty^2)}{4r_\infty^4}.
\end{eqnarray}
It should be noted that the coefficient of $\tilde\sigma_3^2$ approaches a constant. 
Hence the spacetime is asymptotically locally flat, i.e., the asymptotic form of the metric is a twisted $\rm S^1$ bundle over four-dimensional Minkowski spacetime.

The Komar mass at the infinity, the angular momenta associated with the Killing vector fields $\partial_\phi$ and $\partial_\psi$ at the infinity, and the electric charge can be obtained as 
\begin{eqnarray}
&&M_{\rm Komar}=\pi \frac{2r_{\infty }^6(mr_{\infty }^2-q^2)-2a^4(m+q)q^2-a^2(q^4-4mq^2r_{\infty }^2+(4m^2+4mq+3q^2)r_{\infty }^4)}{2r_{\infty }^2(r_{\infty }^6-a^2(q^2-2(m+q)r_{\infty }^2))\rho _{0}}L,\nonumber\\
&&J_{\phi }=0, \\
&&J_{\psi }=-\pi \frac{a(a^2q^3+3q^2r^4_{\infty }-2(2m+q)r^6_{\infty })}{4r^4_{\infty }\sqrt{r^6_{\infty }-a^2(q^2-2(m+q)r^2_{\infty })}}L,\nonumber\\
&&Q=-\frac{\sqrt{3}}{2}\pi q.
\end{eqnarray}

\subsection{Near-Horizon geometry and regularity}
As is seen later, the solution within the region of the parameters in FIG.\ref{fig:region} has two horizons, an event horizon at $r=r_{+}>0$ and an inner horizon at $r=r_{-}>0$, which are determined by the equation $w(r)=0$, where $r_\pm$ are given by
\begin{eqnarray}
r_\pm^2=m\pm \sqrt{(m+q)(m-q-2a^2)}.\label{eq:rpm}
\end{eqnarray}
In the coordinate system $(t,r,\theta,\phi,\psi)$,
 the metric diverges at $r=r_{+}$ and $r=r_{-}$ but these are apparent.
In order to remove this apparent divergence, we introduce the coordinates $(t',\psi')$ such that
\begin{eqnarray}
&&dt=dt'+\frac{\sqrt{h(r)}k(r)}{w(r)}dr,\\
&&d\psi=d\psi'-\frac{\sqrt{h(r)}f(r)k(r)}{w(r)}dr.
\end{eqnarray}
Then the metric takes the form of
\begin{eqnarray}
ds^2=-\frac{w(r)}{h(r)}dt^{\prime 2}-\frac{2k(r)}{\sqrt{h(r)}}dt'dr+\frac{r^2}{4}\biggl[k(r)(\sigma_1^2+\sigma_2^2)+h(r)\left(f(r)dt'+\sigma'_3 \right)^2\biggr].
\end{eqnarray}
This metric well behaves at $r=r_\pm$, i.e., it is smooth at this place.

 Moreover, define the coordinates $(v,\psi'')$ given by
\begin{eqnarray}
&&v=t,\\
&&\psi''=\psi'+f(r_{+})t,
\end{eqnarray}
and then the metric can be rewritten as follows
\begin{eqnarray}
ds^2&=&-\left[\frac{w(r)}{h(r)}+\frac{r^2}{4}h(r)(f(r)-f(r_{+}))^2\right]dv^2-\frac{2k(r)}{\sqrt{h(r)}}dvdr\nonumber\\
&&+\frac{r^2}{2}h(r)(f(r)-f(r_{+}))\sigma''_{3}dt+\frac{r^2}{4}\biggl[k(r)(\sigma^2_1+\sigma ^2_2)+h(r)\sigma_3^{''2}\biggr].\label{eq:metric}
\end{eqnarray}
The Killing vector field $V=\partial_v$ becomes null at $r=r_{+}$ and furthermore, $V$ is hypersurface orthogonal from $V_\mu\propto dr$ there. 

These mean that the hypersurfaces $r=r_{+}$ is a Killing horizon. Similarly, $r=r_-$ is also a Killing horizon. We should also note that in the coordinate system $(v,\phi,\psi'',r,\theta)$, each component of the metric form (\ref{eq:metric}) is analytic on and outside the black hole horizon. Hence the spacetime has no curvature singularity on and outside the black hole horizon.

\begin{figure}[h]
  \begin{center}
\includegraphics[width=0.6\linewidth]{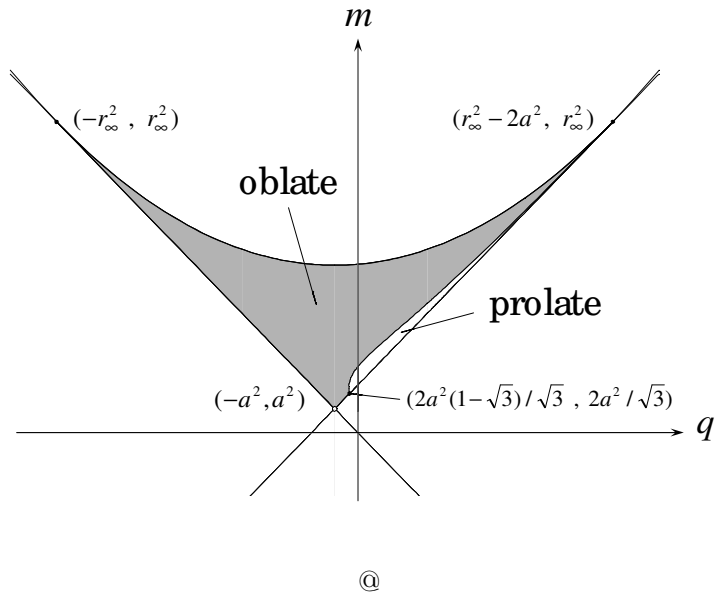}
  \end{center}
@\caption{The aspect ratio of the outer horizon.}
  \label{fig:05}
 \end{figure}

The surface gravity $\kappa$ of the black hole is obtained as
\begin{equation}
\kappa=\frac{r^2_{+}-r^2_{-}}{r_{\infty}(r^2_{\infty}-r^2_{-})}\sqrt{\frac{(r^2_{\infty}-r^2_{+})}{(r^2_{\infty}-r^2_{-})}}\sqrt{\frac{r^6_{\infty}-a^2(q^2-2(m+q)r^2_{\infty})}{r^6_{+}-a^2(q^2-2(m+q)r^2_{+})}}. 
\label{eq:k}
\end{equation} 
Next, we investigate the shape of the horizon, especially, the aspect ratio of $\rm S^2$ base space to the $\rm S^1$ fiber, which characterize the squashing of $\rm S^3$ and is denoted by $k(r_{+})/h(r_{+})$. 
In the case of $k(r_{+})/h(r_{+})>1$, the event horizon is called {\it oblate}, where the radius of $\rm S^2$ larger than that of $\rm S^1$.
In the case of $k(r_{+})/h(r_{+})<1$, the event horizon is called {\it prolate}, where the radius of $\rm S^2$ smaller than that of $\rm S^1$. FIG.\ref{fig:05} shows the oblate region and the prolate region in the $(m,q)$-plane. The shaded region and unshaded region are the oblate region and the prolate region, respectively.
At the boundary of two regions, the ratio is $k(r_{+})/h(r_{+})=1$, where the horizon becomes a round $\rm S^3$. Thus unlike the static solution~\cite{IM} obtained by two of the present authors, the horizon admits a prolate shape in addition to a round $\rm S^3$.

\subsection{Parameters region and absence of CTCs}
We consider the condition that the spacetime has non-degenerate horizons and no CTCs outside the horizons.

\begin{figure}[htbp]
\begin{center}
\includegraphics[width=0.6\linewidth]{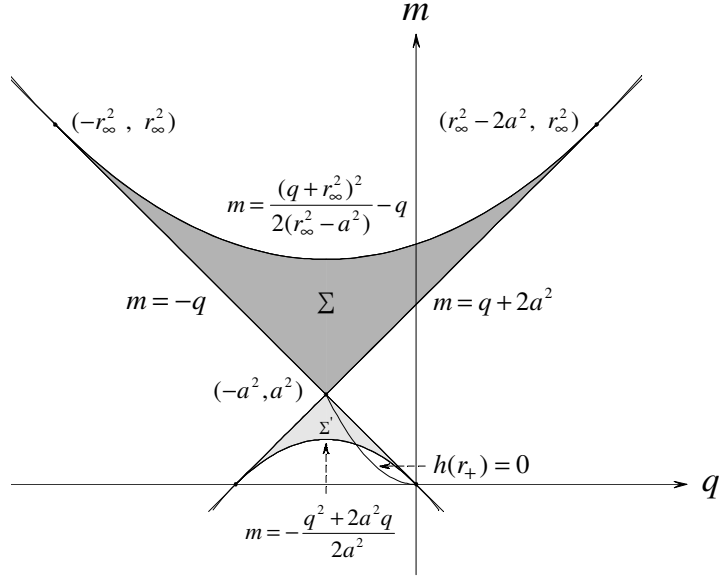}
\end{center}
\caption{
In the region $\Sigma$, since $h(r_{+})>0$, 
the solution has no CTC everywhere outside the horizon, 
but in the region $\Sigma'$, since $h(r_{+})\leq 0$, 
it has CTCs outside the horizon.
The curve which enter only the region $\Sigma'$ and has endpoints at 
the points $(0,0)$ and $(-a^2,a^2)$ denotes $h(r_{+})=0$. 
}
\label{fig:ryouiki3}
\end{figure}

FIG.\ref{fig:region} shows the region of the parameters in a $(q,m)$-plane. The horizons appear at $r>0$ satisfying $w(r)=0$, i.e., the quadratic equation with respect to $r^2$
\begin{eqnarray}
(r^2+q)^2-2(m+q)(r^2-a^2)=0.
\end{eqnarray} 
The necessary and sufficient condition that this quadratic equation has two different roots within the range of $0<r^2<r_\infty^2$ is
\begin{eqnarray}
&&m>0,\label{eq:in2}\\
&&q^2+2(m+q)a^2>0,\label{eq:in3}\\
&&(r_\infty^2+q)^2-2(m+q)(r_\infty^2-a^2)>0,\label{eq:in3-2}\\
&&(m+q)(m-q-2a^2)>0.\label{eq:in4}
\end{eqnarray}
As shown in FIG.\ref{fig:ryouiki3}, the region satisfying the inequalities (\ref{eq:in2})-(\ref{eq:in4}) consist of
 two regions: $\Sigma=\{(q,m)|\ m-q>2a^2,\ m+q>0,\ m<-q+(q+r_\infty^2)^2/(2(r_\infty^2-a^2))\}$ and $\Sigma'=\{(q,m)|\ m-q<2a^2,\ m+q<0,\  m>-(q^2+2a^2q)/(2a^2)\}$. However, as will be seen below, one of the regions $\Sigma'$ is excluded from the requirement for the absence of CTCs outside the black hole horizon.

The positivity of the components $g_{\phi\phi}$ and $g_{\psi\psi}$ outside the horizons assures the absence of CTCs there. Since $g_{\phi\phi}$ is always positive definite outside the horizons, it is sufficient to consider the parameter region such that 
\begin{eqnarray}
g_{\psi\psi}=h(r)=\frac{r^6+2a^2(m+q)r^2-a^2q^2}{r^6}>0
\end{eqnarray}
outside the horizons.
It is noted that the function $h(r)$ is a monotony increasing function of $r$.  
Hence, as a result, the condition for the absence of CTC outside the outer horizons is $h(r_{+})>0$. 
The curve in FIG.\ref{fig:ryouiki3} which enter only the region $\Sigma'$ and has endpoints at the points $(0,0)$ and $(-a^2,a^2)$ denotes $h(r_{+})=0$. 
Since the function $h(r_{+})$ takes a maximum value of a zero in the region $\Sigma'$, it becomes non positive throughout the region $\Sigma'$. In contrast, it takes positive values in the region $\Sigma$. 
This is why in addition to the inequalities (\ref{eq:in2})-(\ref{eq:in4}),  we impose the inequality $m+q>0$ on the parameters.

\subsection{Ergo region} 
In our solutions, an ergo surface is located at $r=r_e (r_+<r_e<r_\infty)$ satisfying the cubic equation with respect to $r^2$
\begin{eqnarray}
-A^2r^6+4r^4-2(aA-2)\left((aA-2)(m+q)+2q\right)r^2+(aA-2)^2q^2=0,
\end{eqnarray}
where the constant $A$ is defined by
\begin{eqnarray}
A=-\frac{2a(q^2-(2m+q)r_\infty^2)}{r_\infty^6-a^2(q^2-2(m+q)r_\infty^2)}.
\end{eqnarray}
In fact, for $r_+^2<r^2<r_\infty^2$, this equation has a single positive root within the region of parameters $\Sigma$, the boundaries $\partial \Sigma_-=\{(q,m)\ |\ m+q=0,-\ r_\infty^2\le q<-a^2\}$ and  $\partial \Sigma_+=\{(q,m)\ |\ m-q-2a^2=0,\ -a^2<q\le r_\infty^2-2a^2\}$.
Thus, there is always an ergo region within the region of parameters (\ref{eq:0in2})-(\ref{eq:0in1}) and $\partial\Sigma_\pm$.

\section{Various limits}\label{sec:limit}

\subsubsection{$r_\infty\to\infty$}
In the limit of $r_\infty\to\infty$ with the other parameters fixed, where the size of an extra dimension becomes infinite, the function $k(r)$ takes the limit of $k(r)\to 1$. Then the metric coincides with that of the asymptotically flat solutions obtained by Cvetic. et.al.~\cite{CLP}. In the specified case of $q\to 0$, the solution reduces to the five-dimensional Myers-Perry black hole solution~\cite{Myers} with two equal angular momenta.

\subsubsection{$q\to 0$}
In the limit of $q\to0$, the solution coincides with the one obtained by Gibbons et.al.~\cite{DM,GW}. This solutions  has only a angular momentum in the direction of an extra dimension. As mentioned previously, Wang regenerated it~\cite{Wang} via the squashing transformation for the five-dimensional Myers-Perry black hole solutions.

\subsubsection{$m\to -q$}
Taking the limit of $m\to -q$ with introducing new coordinates $(\tilde t,\tilde r)$ and the parameters $(\tilde Q,R_{\infty},\tilde J)$ defined as
\begin{eqnarray}
&&r^2=\frac{4(R_{\infty}^2 \tilde r+R_{\infty} \tilde Q)}{\tilde r+R_{\infty}},\\
&&t=\frac{R_{\infty}^2}{R_{\infty}^2-\tilde Q}\tilde t,\\
&&r_\infty^2=4R_{\infty}^2,\\
&&q=-4\tilde Q,\\
&&a=-\frac{2\tilde J}{\tilde Q},
\end{eqnarray}
we obtain the following metric
\begin{eqnarray}
ds^2=-H^{-2}\Bigl[d\tilde t+\frac{\tilde J}{R_{\infty}^2}H_k\sigma_3 \Bigr]^2+Hds_{\rm T-NUT}^2,
\end{eqnarray}
where $ds_{\rm T-NUT}^2$ is the metric on the Euclidean self-dual Taub-NUT space written in the Gibbons-Hawking coordinate and is given by
\begin{eqnarray}
&&ds_{\rm T-NUT}^2=H_k(d\tilde r ^2+\tilde r ^2d\Omega_{S^2}^2)+R_{\infty}^2H_k^{-1}\sigma_3^2
\end{eqnarray}
$H$ and $H_k$ are harmonic functions on the three-dimensional Euclid space and are expressed in the coordinate system as follows
\begin{eqnarray}
&&H=1+\frac{\tilde Q}{R_{\infty}\tilde r},\\
&&H_k=1+\frac{R_{\infty}}{\tilde r}.
\end{eqnarray}
This coincides with the metric of the supersymmetric black hole solutions with a compactified extra dimension on the Euclidean self-dual Taub-NUT space in Ref~\cite{Gaiotto}. This spacetime has an ergo region. 
Moreover, in the case of $r_\infty\to \infty$, the metric can be written as follows
\begin{eqnarray}
ds^2&=&-\left(1+\frac{q}{R^2}\right)^{-2}\left[dt-\frac{aq}{2R^2}\sigma_3\right]^2+\left(1+\frac{q}{R^2}\right)\left[dR^2+R^2d\Omega_{S^3}^2\right],
\end{eqnarray}
where $R^2=r^2+q$ and $d\Omega_{S^3}^2$ is the metric on a unit three-sphere. This is the metric of the BMPV black hole solutions written in terms of the Gibbons-Hawking coordinate.

\subsubsection{$a \to 0$}

The case of $a\to0$ corresponds to the metric of the static non-BPS Kaluza-Klein black hole solution with a squashed horizon obtained by two of authors~\cite{IM}.
\begin{figure}[ht]
  \begin{center}
\includegraphics[width=0.6\linewidth]{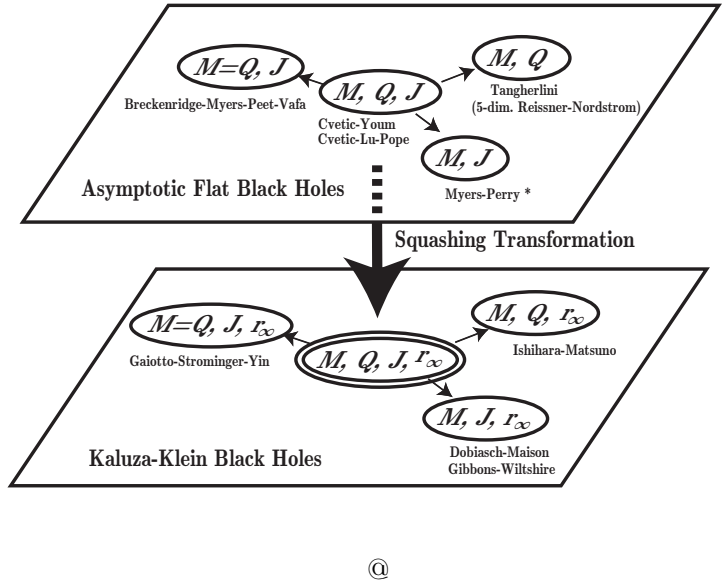}
  \end{center}
@\caption{Various limits: This figure shows the relation between black hole 
solutions 
with asymptotically flatness and Kaluza-Klein black holes. 
* We consider, here, the Myers-Perry solution in the special case of 
two equal angular momenta. 
}
  \label{fig:06}
 \end{figure}

\subsubsection{$m\to q+2a^2$}
In the case of $m\to q+2a^2$, two horizons degenerate, although this is not a BPS solution. In this case, the metric can be written in the form
\begin{eqnarray}
ds^2=-\frac{\Gamma}{r^4}\left(dt+\frac{\omega }{2\Gamma}\sigma_3\right)^2+\frac{r^4}{\Delta } k^2dr^2+\frac{r^2}{4}\left[k(\sigma _1^2+\sigma_2^2)+\frac{\Delta}{\Gamma}\sigma_3^2\right],
\end{eqnarray}
where the functions $\Delta,\ \Gamma,\ \omega $ and $k(r)$ in the metric are given by
\begin{eqnarray}
&&\Delta=(r^2-(q+2a^2))^2,\\
&&\Gamma=r^4-2(q+2a^2)r^2+q^2,\\
&&\omega =a((3q+4a^2)r^2-q^2),\\
&&k(r)=\frac{\left(r_\infty^2-(q+2a^2)\right)^2}{(r_\infty^2-r^2)^2}.
\end{eqnarray}
This spacetime has an ergo region. It is noted that from Eq.(\ref{eq:k}), the surface gravity of the black hole vanishes.

\subsubsection{$\rho_0\to 0$ with $\rho_\pm$ finite}
Here we consider the limit of $\rho_0\to 0$, where the function $k(r)\to 0$.
We introduce the new parameters $\rho _{\pm }$ defined as
\begin{equation}
\rho _{\pm }=\frac{r_{\pm}^2}{r_{\infty}^2-r_\pm^2}\rho _{0}.\label{eq:rho00}
\end{equation}
In terms of $r_\pm$, the constant $\rho_0$ defined in Eq.(\ref{eq:rho0}) is written in the form
\begin{eqnarray}
&&\rho ^2_{0}=\frac{(r^2_{\infty}-r^2_{+})(r^2_{\infty}-r^2_{-})}{4r^2_{\infty}}.
\end{eqnarray}
Then, the metric can rewritten and as
\begin{eqnarray}
&&ds^2=-\bar VdT^2+Ud\rho ^2+\rho (\rho +\rho _{0})d\Omega _{S^2}^2+W\sigma _{3}^2+2HdT\sigma _{3},
\end{eqnarray}
where $T=2\rho _{0}t/r_{\infty}$, and the metric functions $\bar V,U,W$, $H$ in Eqs.(\ref{eq:f1})-(\ref{eq:f2}) are rewritten in the form
\begin{eqnarray}
&&\bar V=\left(1-\frac{\rho _{+}}{\rho}\right)\left(1-\frac{\rho _{-}}{\rho}\right)-\frac{m+q}{2r_{\infty}^2}\left(\frac{a}{\rho_{0}}\right)^2\left(1+\frac{\rho _{0}}{\rho}\right)^2, \\
&&U=\frac{\left(1+\frac{\rho _{0}}{\rho}\right)}{\left(1-\frac{\rho _{+}}{\rho}\right)\left(1-\frac{\rho _{-}}{\rho}\right)}, \\
&&W=\frac{r^6_{\infty}\rho ^3-a^2(\rho +\rho _{0})^2(q^2(\rho +\rho _{0})-2(m+q)r^2_{\infty}\rho )}{4r^4_{\infty}\rho^2(\rho +\rho _{0})}, \\
&&H=-\frac{(\rho +\rho _{0})(r_{\infty}^2(2m+q)\rho-q^2(\rho +\rho _{0})}{4r^3_{\infty}\rho ^2}\left(\frac{a}{\rho_{0}}\right).
\end{eqnarray}

{}From Eq.(\ref{eq:rho00}), the following equations holds
\begin{eqnarray}
&&r^2_{\infty}-r_{+}^2=\frac{r^2_{\infty}}{\rho _{+}}\rho _{0}, \nonumber\\
&&r^2_{\infty}-r_{-}^2=\frac{r^2_{\infty}}{\rho _{-}}\rho _{0}.\label{eq:mai}
\end{eqnarray}
In order that $\rho_\pm$ are finite in the limit of $\rho_0\to 0$, it is necessary that $r_\pm\to r_\infty$. Therefore, from Eq.(\ref{eq:rpm}), a pair of parameters $(q,m)$ must take either limit of 
\begin{eqnarray}
(q,m)\to (-r^2_{\infty},r^2_{\infty})
\end{eqnarray}
or
\begin{eqnarray}
(q,m)\to (r^2_{\infty}-2a^2,r^2_{\infty}).
\end{eqnarray}
 
Furthermore, from Eqs.(\ref{eq:rpm}) and (\ref{eq:mai}), the parameters $(q,m)$ must behave as 
\begin{eqnarray}
(q,m)\simeq (-r_{\infty}^2+\beta _{1}\rho _{0}+\beta_{2}^{-}\rho_0^2,\ r_\infty^2-\beta_1\rho_0+\beta _{2}^{+}\rho _{0}^2),
\end{eqnarray}
or
\begin{eqnarray}
(q,m)\simeq (r_{\infty}^2-2a^2-\beta _{1}\rho _{0}-\beta_2^{-}\rho_0^2,\ r_\infty^2-\beta_1\rho_0+\beta _{2}^{+}\rho _{0}^2),
\end{eqnarray}
respectively, where the constants $\beta_2^{\pm}$ satisfy $\beta_2^-+\beta_2^+=\beta_2$ and the constants $(\beta_1,\beta_2)$ are given by
\begin{eqnarray}
&&(\beta _{1},\beta _{2})=\left(2(\rho _{+}+\rho _{-}),\frac{2(\rho _{+}-\rho _{-})^2}{4\rho _{+}\rho _{-}-a^2}\right). 
\end{eqnarray}
(i) In the case of $(q,m)\simeq (-r_{\infty}^2+\beta _{1}\rho _{0}+\beta_{2}^{-}\rho_0^2,\ r_\infty^2-\beta_1\rho_0+\beta _{2}^{+}\rho _{0}^2)$, the metric becomes
\begin{eqnarray}
&& ds^2=-\frac{4(\rho -\rho _{+})(\rho -\rho _{-})-a^2}{4\rho ^2}dT^{2} \nonumber +\frac{d\rho ^2}{\left(1-\frac{\rho _{+}}{\rho}\right)\left(1-\frac{\rho _{-}}{\rho}\right)}\nonumber\\
 &&+\rho ^2d\Omega ^2_{S^2}+\frac{4\rho_{+}\rho_{-}-a^2}{4}\sigma ^{2}_{3}+a\frac{\sqrt{4\rho _{+}\rho _{-}-a^2}}{2\rho }dT\sigma_{3},\label{limit7}
\end{eqnarray}
where the coordinates $\psi$ and $T$ are transformed as
\begin{eqnarray}
&&\psi\to \psi -\frac{a(\rho _{+}-\rho _{-})}{\sqrt{\rho _{+}\rho _{-}}(4\rho _{+}\rho _{-}-a^2)}\ T, \\
&&T\to \sqrt{\frac{4\rho _{+}\rho _{-}}{4\rho _{+}\rho _{-}-a^2}}\ T.  
\end{eqnarray}
A coefficient of $\sigma^{2}_3$, i.e., the size of the $\rm S^1$ fiber, takes the constant value.

(ii) In the case of $(q,m)\simeq (r_{\infty}^2-2a^2-\beta _{1}\rho _{0}-\beta_2^{-}\rho_0^2,\ r_\infty^2-\beta_1\rho_0+\beta _{2}^{+}\rho _{0}^2)$, the metric reduces to
\begin{eqnarray}
&& ds^2=-\frac{16\rho^2_{+}\rho^2_{-}(\rho -\rho _{+})(\rho -\rho _{-})-a^2(a^2-6\rho_{+}\rho_{-})^2}{16\rho^2_{+}\rho^2_{-}\rho ^2}dT^{2} \nonumber \\
&&+\frac{d\rho ^2}{\left(1-\frac{\rho _{+}}{\rho}\right)\left(1-\frac{\rho _{-}}{\rho}\right)}+\rho ^2d\Omega ^2_{S^2}+\frac{(4\rho_{+}\rho_{-}-a^2)(2\rho_{+}\rho_{-}+a^2)^2}{16\rho ^2_{+}\rho ^2_{-}}\sigma^{2} _{3} \nonumber \\
&&\hspace{1cm}+a\frac{a^2-6\rho_{+}\rho_{-}}{2\rho_{+}\rho_{-}\rho }\sqrt{\frac{(4\rho _{+}\rho _{-}-a^2)(2\rho _{+}\rho _{-}+a^2)^2}{16\rho^2 _{+}\rho^2 _{-}}} dT\sigma^{} _{3}  \label{limit8}
\end{eqnarray}
where the coordinates $\psi$ and $T$ are transformed as
\begin{eqnarray}
&&\psi\to \psi+ \frac{r^2_{\infty}(2m+q)+q^2}{r^6_{\infty}-a^2(q^2-2(m+q)r^2_{\infty}}\frac{ar_{\infty}}{\rho_0} \ T\\
&&T\to \sqrt{\frac{16\rho^3 _{+}\rho^3 _{-}}{(4\rho _{+}\rho _{-}-a^2)(2\rho _{+}\rho _{-}+a^2)^2}}\ T.
\end{eqnarray}
The size of the $S^1$ fiber in this limit also takes the constant value.

When $a=0$, both of two limits become four-dimensional Reissner-Nordstr\"{o}m black hole with a twisted $S^1$ bundle, where the size of the $\rm S^1$
fiber takes the constant value $r_{\infty}/2$ ~\cite{IM}.

Finally, introducing the coordinate $\psi\to\sqrt{4\rho _{+}\rho _{-}-a^2}\psi/2 $ in Eq.(\ref{limit7}), or $\psi\to 3\sqrt{4\rho _{+}\rho _{-}-a^2}\psi/2$ in Eq.(\ref{limit8}), and taking the limit of $a^2\to 4\rho_+\rho_-$ , we find both metric are reduced to
\begin{eqnarray}
&&ds^2=-\left(1-\frac{\rho_{+}+\rho_{-}}{\rho }\right)dt^2+\frac{d\rho ^2}{\left(1-\frac{\rho _{+}}{\rho}\right)\left(1-\frac{\rho _{-}}{\rho}\right)}+\rho ^2d\Omega ^2_{S^2}+d\psi ^2+\frac{a}{\rho }dtd\psi. 
\end{eqnarray}
Moreover, transforming the coordinates as
$t \to-\sqrt{\frac{\rho_{+}}{\rho_{+}-\rho_{-}}}t-\sqrt{\frac{\rho_{-}}{\rho_{+}-\rho_{-}}}\psi$ and $\psi\to -\sqrt{\frac{\rho_{-}}{\rho_{+}-\rho_{-}}}t-\sqrt{\frac{\rho_{+}}{\rho_{+}-\rho_{-}}}\psi$,
we obtain the the metric and the gauge potential: 
\begin{eqnarray}
ds^2=-\left(1-\frac{\rho_{+}}{\rho }\right)dt^{2}+\frac{d\rho ^2}{\left(1-\frac{\rho _{+}}{\rho}\right)\left(1-\frac{\rho _{-}}{\rho}\right)}+\rho ^2d\Omega ^2_{S^2}+\left(1-\frac{\rho_{-}}{\rho }\right)d\psi^{2},
\end{eqnarray}
\begin{eqnarray}
\textbf{A}=\frac{\sqrt{3}}{2}\sqrt{\rho_{+}\rho_{-}}\cos\theta d\phi. 
\end{eqnarray}
This coincides with the metric of the magnetically charged black string solution in Ref.\cite{HS}.

\section{Summary}\label{sec:summary}
We have constructed a new rotating charged Kaluza-Klein black hole solution in the five-dimensional Einstein-Maxwell theory with a Chern-Simon term. The features of the solutions have been also investigated. The spacetime is asymptotically locally flat, i.e., a twisted $\rm S^1$ bundle over the four-dimensional Minkowski spacetime. This solution has four parameters, the mass, the angular momenta in the direction of an extra dimension, the electric charge and the size of the extra dimension. The solution describes the physical situation such that in general a non-BPS black hole is boosted in the direction of the extra dimension. As the interesting feature of the solution, unlike the static previous solution~\cite{IM}, the horizon admits a prolate shape in addition to a round $\rm S^3$.  The solution has the limits to the supersymmetric black hole solution and a new extreme non-BPS black hole solutions and a new rotating black hole solution with a constant twisted $\rm S^1$ fiber.

\section*{Acknowledgments}
This work is supported by the Grant-in-Aid
for Scientific Research No.19540305.

\end{document}